\documentclass[aps,prl,twocolumn,preprintnumbers,showpacs,reprint,nofootinbib]{revtex4-1}
\usepackage{amsmath,amssymb}
\usepackage{latexsym}
\usepackage{epsfig}
\usepackage[english]{babel}

\newcommand{\eq}{\begin{equation}}
\newcommand{\feq}{\end{equation}}
\newcommand{\eqn}{\begin{eqnarray}}
\newcommand{\feqn}{\end{eqnarray}}
\newcommand{\arr}{\begin{eqnarray*}}
\newcommand{\farr}{\end{eqnarray*}}

\font\mybb=msbm10 at 12pt
\def\bb#1{\hbox{\mybb#1}}

\def\bR {\bb{R}}

\begin{document}

\title{Four-dimensional black holes with unusual horizons}

\author{Dietmar Klemm}
\affiliation{Dipartimento di Fisica, Universit\`a di Milano, and \\
INFN, Sezione di Milano, \\
Via Celoria 16, I-20133 Milano.}

\preprint{IFUM-1022-FT}

\begin{abstract}

We show that in presence of a cosmological constant or, more generally, of a
scalar potential, there can exist actually more possibilities for the horizon geometry of a
four-dimensional black hole than the hitherto known spherical, hyperbolic or flat cases.
In particular, there are black holes whose event horizons are noncompact manifolds with
finite volume, which are topologically spheres with two punctures. We give concrete
examples of such black holes in Einstein-Maxwell-AdS gravity and discuss their thermodynamics.
These exotic solutions,
that seem to have been overlooked in the existing literature, may provide interesting testgrounds
to address questions related to black hole physics or holography.

\end{abstract}


\maketitle

\section{Introduction}
\label{intro}

More than forty years ago Hawking proved his famous theorem \cite{Hawking:1971vc,Hawking:1973uf}
on the topology of black holes, which asserts that event horizon cross sections of 4-dimensional
asymptotically flat stationary black holes obeying the dominant energy condition are topologically
$\text{S}^2$. This result extends to outer apparent horizons in black hole spacetimes that are
not necessarily stationary \cite{Hawking:72}. Such restrictive uniqueness theorems do not hold in higher
dimensions, the most famous counterexample being the black ring of Emparan and Reall
\cite{Emparan:2001wn}, with horizon topology $\text{S}^2\times\text{S}^1$.
Nevertheless, Galloway and Schoen \cite{Galloway:2005mf} were able to show that, in arbitrary
dimension, cross sections of the event horizon (in the stationary case) and outer apparent horizons
(in the general case) are of positive Yamabe type, i.e., admit metrics of positive scalar curvature.

In four dimensions, one can have black holes with nonspherical horizons by relaxing some of the
assumptions that go into Hawking's theorem. For instance, in asymptotically anti-de~Sitter (AdS) space,
the horizon of a black hole can be a compact Riemann surface $\Sigma_g$ of any genus
$g$ \cite{Lemos:1994xp}. In this case, both the asymptotically flat and dominant energy conditions are
violated. It should be noted that, unless $g=0$, these spacetimes are asymptotically only locally
AdS; their global structure is different. This is in contrast to the black rings in five dimensions,
which are asymptotically Minkowski, in spite of their nontrivial horizon topology.
It is also possible to add rotation to these black hole solutions in $\text{AdS}_4$ \cite{Klemm:1997ea},
but in this case the horizon cannot be compactified anymore, and the resulting spacetimes
describe rotating black branes\footnote{Among the solutions found in \cite{Klemm:1997ea} there
is a rotating cylindrical black hole. A spinning toroidal solution should in principle exist, but has not
yet been constructed.}.

The aim of this letter is to show that these possibilities do not exhaust the spectrum of potential
horizon geometries of asymptotically $\text{AdS}_4$ black holes. In particular, we will see that
there exist black holes whose event horizons are noncompact manifolds with yet finite volume
(and thus finite entropy), which are topologically spheres with two punctures.
Solutions of this type were presented for the first time (and briefly discussed) in \cite{Gnecchi:2013mja}
in the context of $N=2$ gauged supergravity coupled to vector multiplets. Here we will give a more
detailed description of their counterparts in Einstein-Maxwell-$\Lambda$ gravity, with emphasis
on their thermodynamics and global structure. These black holes represent in some sense (that will be
explained below) the ultraspinning limit of the Kerr-Newman-AdS solution, when the
rotation parameter $j$ approaches
the AdS curvature radius $l$. This limit cannot be taken directly in the KNAdS metric, since this becomes
singular for $j\to l$. However, the authors of \cite{Caldarelli:2008pz,Caldarelli:2012cm} showed that one
can take a finite limit by zooming into the pole, and we will prove here that this limit coincides
precisely with our solution close to the punctures of the sphere. It turns out that, in terms of the mass
$M$ and angular momentum $J$, the ultraspinning property translates into the chirality condition
$M=-J/l^2$. This means that these exotic black holes are described by chiral excitations of a
three-dimensional conformal field theory.

As we said, we study here in detail only the Einstein-Maxwell-$\Lambda$ case.
A more extensive discussion of the corresponding solutions in $N=2$ matter-coupled
gauged supergravity constructed in \cite{Gnecchi:2013mja} will be presented elsewhere.

\section{Noncompact horizons with finite volume}

Let us start with the Carter-Pleba\'nski solution \cite{Carter:1968ks,Plebanski:1975} of
Einstein-Maxwell-Lambda theory, whose metric and $\text{U}(1)$ field strength are respectively
given by
\begin{eqnarray}
ds^2 &=& -\frac{Q(q)}{p^2+q^2}(d\tau - p^2d\sigma)^2 + \frac{p^2+q^2}{Q(q)}dq^2 \nonumber \\
         &&\quad    + \frac{p^2+q^2}{P(p)}dp^2 + \frac{P(p)}{p^2+q^2}(d\tau + q^2d\sigma)^2\,, 
                      \label{metr-CP}
\end{eqnarray}
\begin{eqnarray}
F &=& \frac{\mathsf{Q}(p^2-q^2)+2\mathsf{P}pq}{(p^2+q^2)^2}dq\wedge (d\tau - p^2d\sigma)
          \nonumber \\
&&\quad + \frac{\mathsf{P}(p^2-q^2)-2\mathsf{Q}pq}{(p^2+q^2)^2}dp\wedge (d\tau + q^2d\sigma)\,,
         \label{F-CP}
\end{eqnarray}
where the quartic structure functions read
\begin{eqnarray}
P(p) &=& \alpha - \mathsf{P}^2 + 2np - \varepsilon p^2  + (-\Lambda/3)p^4\,, \nonumber \\
Q(q) &=& \alpha + \mathsf{Q}^2 - 2mq + \varepsilon q^2 + (-\Lambda/3)q^4\,. \label{struc-func}
\end{eqnarray}
Here, $\mathsf{Q}$, $\mathsf{P}$ and $n$ denote the electric, magnetic and NUT-charge respectively, $m$
is the mass parameter, while $\alpha$ and $\varepsilon$ are additional non-dynamical
constants. In what follows, we shall consider only the case of a negative cosmological constant,
$\Lambda=-3/l^2$.

\subsection{Global structure and horizons}

Depending on the number of real roots of the polynomial $P(p)$, the following scenarios
are possible \cite{Gnecchi:2013mja}: If there are four distinct roots $p_a<p_b<p_c<p_d$, and we
consider the region $p_b\le p \le p_c$ (where $P\ge 0$), we get black holes with
spherical horizon topology. Restricting instead to the other regions where $P$ is positive,
namely $p\ge p_d$ or $p\le p_a$, leads to the rotating hyperbolic black holes first discovered in
\cite{Klemm:1997ea}. If $P$ has no real roots, one obtains rotating generalizations of the AdS
black brane \cite{Klemm:1997ea}. An interesting situation occurs if two or more roots coincide.
Let us consider this case for vanishing NUT charge, $n=0$. Then $P$ has two double roots
at $p=\pm p_a$, where $p_a^2=\varepsilon l^2/2$, if the parameters are constrained by
\eq
l^2\varepsilon^2 = 4(\alpha - \mathsf{P}^2)\,. \label{constraint-par}
\feq
By using the scaling symmetry
\begin{eqnarray}
p &&\to \lambda p\,, \quad q \to \lambda q\,, \quad \tau \to \tau/\lambda\,, \quad \sigma \to
\sigma/\lambda^3\,, \nonumber \\
\alpha &&\to \lambda^4\alpha\,, \quad \mathsf{P} \to \lambda^2\mathsf{P}\,, \quad
\mathsf{Q} \to \lambda^2\mathsf{Q}\,, \nonumber \\
\quad m &&\to \lambda^3 m\,, \quad n \to \lambda^3 n\,, \quad \varepsilon \to
\lambda^2\varepsilon\,, \label{scaling-symm}
\end{eqnarray}
that leaves the solution \eqref{metr-CP}, \eqref{F-CP} invariant, we can set $\varepsilon=2$ and thus
$p_a=l$ without loss of generality. One has then
\eq
Q(q) = \left(l + \frac{q^2}{l}\right)^2 + \mathsf{P}^2 + \mathsf{Q}^2 - 2mq\,, \label{Q(q)}
\feq
whose largest root $q_{\text h}$ yields the location of the horizon. It turns out that there is
a lower bound on the mass parameter $m$ in order for horizons to exist, namely
\eq
m \ge m_0 \equiv 2q_{\text{h},0}\left(\frac{q_{\text{h},0}^2}{l^2} + 1\right)\,,
\feq
where
\eq
q_{\text{h},0}^2 \equiv \frac{l^2}3\left[-1 + \left(4 + \frac{3}{l^2}(\mathsf{P}^2 + \mathsf{Q}^2)
\right)^{1/2}\right]\,.
\feq
For $m=m_0$, $Q$ has a double root at $q=q_{\text{h},0}$, and thus the black hole is extremal.
If $m>m_0$, we have a nonextremal black hole, whereas for $m<m_0$ there is a naked singularity.

Prior to analyzing the horizon geometry, let us consider the asymptotic behaviour for $q\to\infty$.
The metric on the conformal boundary of \eqref{metr-CP} is given by
\eq
ds^2_{\text{bdry}} = -(d\tau + (\mu - p^2)d\sigma)^2 + l^2\left[\frac{dp^2}{P} + P d\sigma^2\right]\,,
\label{metr-bdry1}
\feq
where we took into account a possible shift $\tau\to\tau+\mu\sigma$ of the time coordinate.
If $\sigma$ is a noncompact coordinate, $\mu$ can take any value, but if we want to compactify
$\sigma$ ($\sigma\sim\sigma+L$, which we shall assume in what follows), $\mu$ must be equal
to $l^2$ in order to avoid closed timelike curves (CTCs) on the boundary. In that case,
\eqref{metr-bdry1} boils down to
\eq
ds^2_{\text{bdry}} = -d\tau^2 + 2(p^2-l^2)d\tau d\sigma + l^2\frac{dp^2}{P}\,. \label{metr-bdry2}
\feq
It can be easily shown (using $m\ge m_0$ and $q_{\text h}\ge q_{\text{h},0}$) that for the choice
$\mu=l^2$, $g_{\sigma\sigma}$ is always positive in the region $q_{\text h}\le q<\infty$, so
that there are no CTCs outside the horizon. On the conformal boundary itself, $\sigma$ becomes
a (compact) null coordinate, but this is something we are used to from discrete light cone quantization.

\begin{figure}[htb]
  \begin{center}
    \includegraphics[scale=0.55]{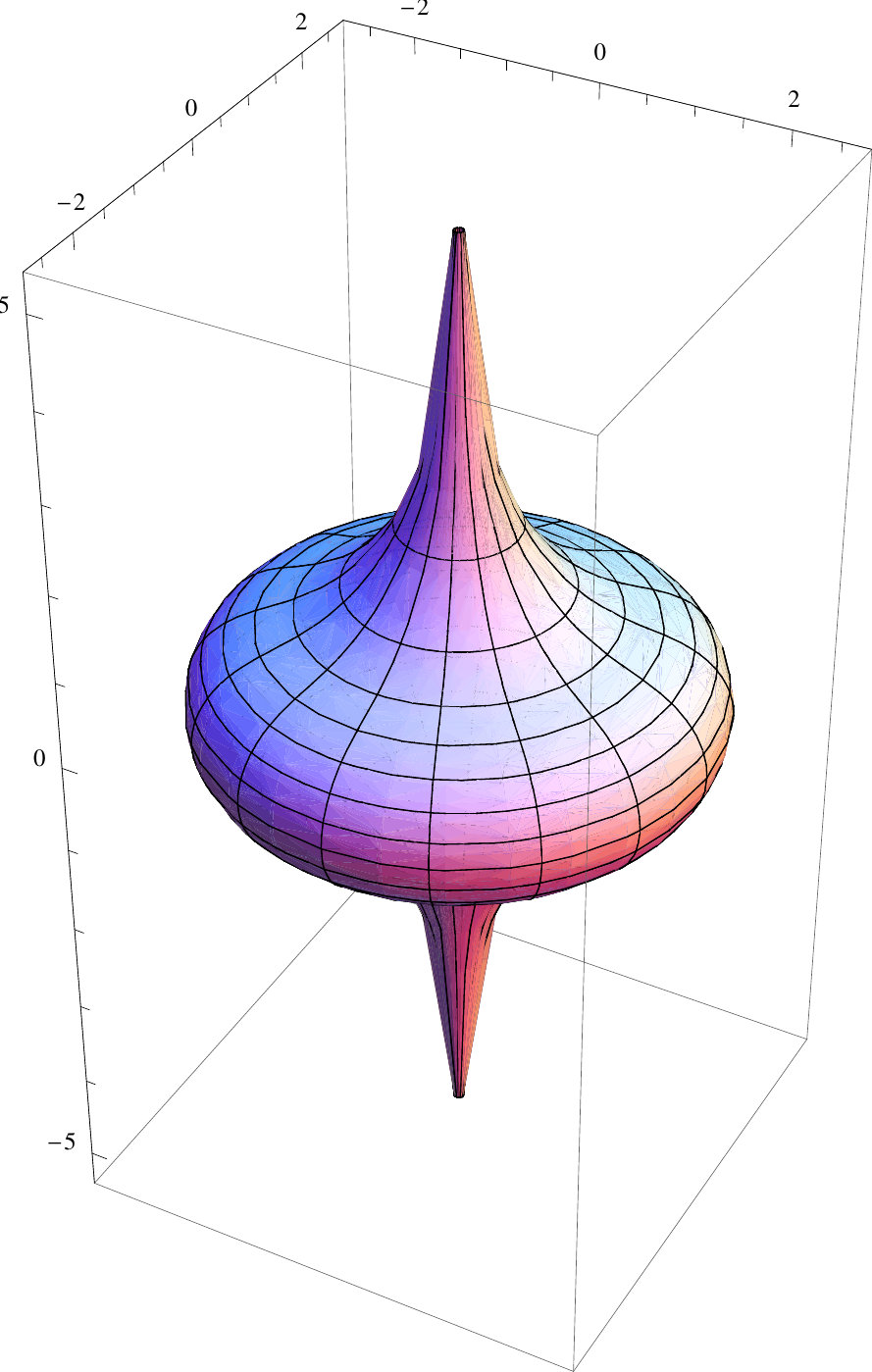}
  \end{center}
  \caption{The event horizon of a black hole in the case where $P(p)$ has two double roots, embedded in
  $\bR^3$ as a surface of revolution.\label{cusp}}
\end{figure}

Let us now take a deeper look at the geometry of the horizon. Its induced metric reads
\eq
ds^2_{\text{hor}} = \frac{p^2+q_{\text h}^2}{P(p)}dp^2 + \frac{P(p)}{p^2 + q_{\text h}^2}
                               (l^2 + q_{\text h}^2)^2 d\sigma^2\,, \label{metr-hor}
\feq
where $P(p)=(p^2-l^2)^2/l^2$, and we consider the range $-l\le p\le l$. Obviously \eqref{metr-hor}
becomes singular for $p^2=l^2$. To understand more in detail what happens at these singularities,
take for instance the limit $p\to l$, in which \eqref{metr-hor} simplifies to
\eq
ds^2_{\text{hor}} = (l^2+q_{\text h}^2)\left[\frac{d\rho^2}{4\rho^2} + 4\rho^2 d\sigma^2\right]\,.
\label{metr-hor-limit}
\feq
Here, the new coordinate $\rho$ is defined by $\rho=l-p$. \eqref{metr-hor-limit} is a metric
of constant negative curvature on the hyperbolic space $\text{H}^2$ (actually on a quotient
thereof, since we chose $\sigma$ to be a compact coordinate). Because \eqref{metr-hor} is
symmetric under $p\to -p$, an identical result holds for $p\to -l$. Thus, for $p\to\pm l$, the
horizon approaches a space of constant negative curvature, and there is no true singularity
there. In particular, this implies that the horizon is noncompact. Nevertheless, the horizon
area
\eq
A_{\text h} = \int (l^2+q_{\text h}^2)d\sigma dp = 2lL(l^2+q_{\text h}^2)
\feq
is finite. We see that, though being noncompact, the event horizon has finite area, and the
entropy
\eq
S = \frac{A_{\text h}}{4G} = \frac{lL}{2G}(l^2+q_{\text h}^2) \label{S}
\feq
is thus also finite. In order to visualize the geometry \eqref{metr-hor}, one can embed it in $\bR^3$
as a surface of revolution, cf.~\cite{Gnecchi:2013mja} for details. The result is shown in figure
\ref{cusp} for the values $l=1$, $L=2\pi$ and $q_{\text h}^2=5$. The two cusps extend up to
infinity, with $p\to\pm l$ for the upper (lower) cusp respectively. The `equator', where the radius of
the surface of revolution becomes maximal, is reached for $p=0$.

In \cite{Gnecchi:2013mja}, it was furthermore shown that the case of two double roots of $P(p)$
considered here corresponds to taking the limit $j\to l$ for the rotation parameter $j$ of the spherical
Kerr-Newman-AdS black hole\footnote{This can be seen as follows: In the generic case, $P(p)$ has
(for $n=0$) 4 distinct roots $\pm p_a,\pm p_b$ with $0<p_a<p_b$. The KNAdS solution is
recovered from \eqref{metr-CP} by considering the region $-p_a\le p\le p_a$, where $P(p)\ge 0$, using
the scaling symmetry \eqref{scaling-symm} to set $p_b=l$, and defining the rotation parameter $j$
by $p_a^2=j^2$ \cite{Gnecchi:2013mja}. Thus, the limit of coincident roots, $p_a=p_b$,
corresponds to $j^2=l^2$.}. This `ultraspinning' limit cannot be taken directly in the KNAdS solution,
since the latter becomes singular for $j=l$. However, the limit $j\to l$ while keeping the horizon size
finite and simultaneously zooming into the pole, is well-defined \cite{Caldarelli:2008pz,Caldarelli:2012cm},
leading in the uncharged case to (cf.~e.g.~(5.25) of \cite{Caldarelli:2012cm})
\begin{eqnarray}
ds^2 &=& -V(r)\left[dt + 4n\sinh^2\!\frac{\theta}2 d\phi\right]^2 + \frac{dr^2}{V(r)} \nonumber \\
         &  & + (r^2+n^2)(d\theta^2 + \sinh^2\!\theta d\phi^2)\,, \label{limit-Marco}
\end{eqnarray}
where $n=l/2$, $V(r)=\Delta_r/(r^2+n^2)$ and
\eq
\Delta_r = n^2 - r^2 + \frac{r^4}{l^2} + \frac{6n^2r^2}{l^2} - \frac{3n^4}{l^2} - \frac{mr}{4}\,.
\feq
This can be compared with the expansion of \eqref{metr-CP} (with $P(p)=(p^2-l^2)/l^2$ and zero
charges) for $p\to l$, which gives (after shifting $\tau\to\tau+l^2\sigma$)
\begin{eqnarray}
ds^2 &=& -\frac{Q(q)}{l^2+q^2}\left[d\tau + 2l\rho d\sigma\right]^2 + \frac{l^2+q^2}{Q(q)}dq^2
                \nonumber \\
         &  & + (l^2+q^2)\left[\frac{d\rho^2}{4\rho^2} + 4\rho^2 d\sigma^2\right]\,, \label{expansion-4d}
\end{eqnarray}
where $\rho=l-p$. By the coordinate transformation $q=2r$,
\begin{displaymath}
\frac{4\rho(\sigma - i) + i}{4\rho(\sigma + i) + i} = e^{i\phi}\tanh\!\frac{\theta}2 \equiv z\,, \quad
2\tau = t + il\ln\frac{z-1}{\bar z-1}\,,
\end{displaymath}
\eqref{expansion-4d} can be cast precisely into \eqref{limit-Marco}. Notice also that the expansion
of the boundary metric \eqref{metr-bdry2} near $p=l$ yields
\eq
ds^2_{\text{bdry}} \to -(d\tau + 2l\rho d\sigma)^2 + l^2\left[\frac{d\rho^2}{4\rho^2} + 4\rho^2
d\sigma^2\right]\,,
\feq
which is nothing else than $\text{AdS}_3$, written as a Hopf-like fibration over $\text{H}^2$.

Before we come to the thermodynamics of these exotic black holes, let us briefly consider other
root configurations of $P(p)$. One might imagine a scenario with two single roots $p_a<p_b$
and a double root $p_c=p_d>p_b$. If we restrict to the region $p_b\le p\le p_c$, where $P(p)$
is positive, this would lead to a black hole with a noncompact horizon that is topologically a sphere
with one puncture instead of two. It is easy to see that this requires
\eq
p_a + p_b + 2p_c = 0\,, \label{constraint}
\feq
plus some other constraints on the roots and parameters. Now consider the limit $p\to p_b$,
where $P\to\kappa(p-p_b)$ for some positive constant $\kappa$. As before, we shift
$\tau\to\tau+\mu\sigma$ in \eqref{metr-CP}, and demand $g_{\sigma\sigma}\ge 0$ to avoid CTCs.
This fixes $\mu=p_b^2$. But then, for $p\to p_c$, $g_{\sigma\sigma}$ will become negative
unless $p_b=-p_c$. Plugging this into \eqref{constraint} yields $p_a=p_b$, which brings us back
to the situation of two double roots studied before. It remains to be seen if in the solutions to
matter-coupled gauged supergravity constructed in \cite{Gnecchi:2013mja,Chow:2013gba}, where
the polynomial $P(p)$ is more general, such a scenario is possible. Notice also that in de~Sitter space,
in principle one might have even more exotic configurations, for instance one single and one triple
root. We shall leave a deeper analysis of these cases for future work.

\subsection{Thermodynamics}

In what follows, we shall discuss only the case of vanishing magnetic charge, $\mathsf{P}=0$.
Writing the metric \eqref{metr-CP} in the ADM form (again after shifting $\tau\to\tau+l^2\sigma$)
\begin{eqnarray}
ds^2 &=& -N^2 d\tau^2 + f(d\sigma - \omega d\tau)^2 \nonumber \\
         &  & \qquad + (p^2+q^2)\left(\frac{dq^2}{Q} + \frac{dp^2}{P}\right)\,,
\end{eqnarray}
where the functions $N,f,\omega$ are not reported here, gives the angular velocity of the horizon
\eq
\omega_{\text h} = \omega|_{q=q_{\text h}} = -\frac1{q_{\text h}^2 + l^2}\,. \label{omega}
\feq
The temperature can be obtained by requiring the absence of conical singularities in the
quasi-Euclidean section $\tau\to-i\tau_{\text E}$, with the result
\eq
T = \frac{Q'(q)|_{q=q_{\text h}}}{4\pi(q_{\text h}^2 + l^2)}\,, \label{T}
\feq
where a prime denotes differentiation w.r.t.~$q$. The electric charge, mass and angular
momentum are respectively
\begin{displaymath}
\mathsf{Q}_{\text{el}} = \frac1{4\pi G}\oint{^{\star\!}F} = \frac{lL\mathsf{Q}}{2\pi G}\,, \quad
M = \frac{lLm}{2\pi G}\,, \quad J = -\frac{l^3Lm}{2\pi G}\,.
\end{displaymath}
$M$ and $J$ were computed as Komar integrals associated to the Killing vectors $\partial_{\tau}$
and $\partial_{\sigma}$ respectively. Note that, in order to compute $M$, we subtracted the
background solution with $m=0$\footnote{Using the results of \cite{AlonsoAlberca:2000cs}
together with the constraint \eqref{constraint-par}, it is straightforward to shew that this background
with $m=n=\mathsf{P}=0$ is supersymmetric.}.
Notice also that $M$ and $J$ are related by the chirality-type
condition $M=-J/l^2$. This is a consequence of the fact that the case of two double roots of $P(p)$
considered here corresponds to taking the ultraspinning limit $j\to l$ for the rotation parameter $j$ of
the spherical KNAdS black hole. The angular momentum is thus not independent of the mass. In this
situation, it is more convenient to use $L_0=(M+J/l^2)/2$, $\tilde L_0=(M-J/l^2)/2$ in place of $M$, $J$
as thermodynamic variables. For $S=S(L_0,\tilde L_0,\mathsf{Q}_{\text{el}})$ the first law must be
\eq
TdS = (1-\Omega l^2)dL_0 + (1+\Omega l^2)d\tilde L_0 - \phi_{\text{el}} d\mathsf{Q}_{\text{el}}\,,
\label{first-law}
\feq
where $\Omega$ and $\phi_{\text{el}}$ denote the angular velocity and electric potential respectively.
In our case $L_0=0$, so the $dL_0$ term is absent. Using the
equation $Q(q_{\text h})=0$ together with \eqref{S} and the expressions for $M$ and $J$, it is
straightforward to obtain the thermodynamic fundamental relation
\eq
\tilde L_0 = \frac{G}{\pi lL}\left[\left(\frac Sl\right)^2 + \left(\pi\mathsf{Q}_{\text{el}}\right)^2\right]
                   \left(\frac{2SG}{lL} - l^2\right)^{-1/2}\,.
\feq
Using this, one easily checks that \eqref{first-law} indeed holds, with $T$ given by \eqref{T},
$\Omega=\omega_{\text h}$, $L_0=0$, and
\eq
\phi_{\text{el}} = A_{\mu}\chi^{\mu}|_{q=q_{\text h}} = \frac{\mathsf{Q}q_{\text h}}{q_{\text h}^2 + l^2}\,,
\feq
where $\chi=\partial_{\tau}+\omega_{\text h}\partial_{\sigma}$ is the null generator of the
horizon and
\eq
A = \frac{\mathsf{Q}q}{p^2+q^2}(d\tau + (l^2 - p^2)d\sigma)
\feq
the vector potential.

\section{Final remarks}

In this letter, we reported on a new type of black holes whose horizons are noncompact manifolds
with finite volume. We discussed their global structure, relation to the ultraspinning limit of
the KNAdS solution, and thermodynamics. It was shown that the mass and angular momentum
obey the chirality condition $M=-J/l^2$, which indicates that the black hole microstates are
chiral excitations of a three-dimensional conformal field theory. It would be interesting to
elaborate on this point in more detail. A further direction for future research would be an investigation
of the Euclidean version of our solutions, and of its holographic interpretation, similar to what was
done in \cite{Martelli:2013aqa}. We hope to come back to this in the future.


\begin{thebibliography}{99}

%\cite{Hawking:1971vc}
\bibitem{Hawking:1971vc}
  S.~W.~Hawking,
  %``Black holes in general relativity,''
  Commun.\ Math.\ Phys.\  {\bf 25} (1972) 152.
  %%CITATION = CMPHA,25,152;%%

%\cite{Hawking:1973uf}
\bibitem{Hawking:1973uf}
  S.~W.~Hawking and G.~F.~R.~Ellis,
  %``The large scale structure of space-time,''
  Cambridge University Press, Cambridge, 1973

\bibitem{Hawking:72}
  S.~W.~Hawking, in `Black Holes, Les Houches Lectures' (1972), edited by
  C.~DeWitt and B.~S.~DeWitt (North Holland, Amsterdam, 1972).

%\cite{Emparan:2001wn}
\bibitem{Emparan:2001wn}
  R.~Emparan and H.~S.~Reall,
  %``A rotating black ring solution in five-dimensions,''
  Phys.\ Rev.\ Lett.\  {\bf 88} (2002) 101101
  [hep-th/0110260].
  %%CITATION = HEP-TH/0110260;%%

%\cite{Galloway:2005mf}
\bibitem{Galloway:2005mf}
  G.~J.~Galloway and R.~Schoen,
  %``A generalization of Hawking's black hole topology theorem to higher dimensions,''
  Commun.\ Math.\ Phys.\  {\bf 266} (2006) 571
  [gr-qc/0509107].
  %%CITATION = GR-QC/0509107;%%

%\cite{Lemos:1994xp}
\bibitem{Lemos:1994xp}
  J.~P.~S.~Lemos,
  %``Cylindrical black hole in general relativity,''
  Phys.\ Lett.\ B {\bf 353} (1995) 46
  [gr-qc/9404041];
  %%CITATION = GR-QC/9404041;%%
%\cite{Mann:1996gj}
%\bibitem{Mann:1996gj}
  R.~B.~Mann,
  %``Pair production of topological anti-de Sitter black holes,''
  Class.\ Quant.\ Grav.\  {\bf 14} (1997) L109
  [gr-qc/9607071];
  %%CITATION = GR-QC/9607071;%%
%\cite{Vanzo:1997gw}
%\bibitem{Vanzo:1997gw}
  L.~Vanzo,
  %``Black holes with unusual topology,''
  Phys.\ Rev.\ D {\bf 56} (1997) 6475
  [gr-qc/9705004];
  %%CITATION = GR-QC/9705004;%%
%\cite{Cai:1996eg}
%\bibitem{Cai:1996eg}
  R.~-G.~Cai and Y.~-Z.~Zhang,
  %``Black plane solutions in four-dimensional space-times,''
  Phys.\ Rev.\ D {\bf 54} (1996) 4891
  [gr-qc/9609065].
  %%CITATION = GR-QC/9609065;%%

%\cite{Klemm:1997ea}
\bibitem{Klemm:1997ea}
  D.~Klemm, V.~Moretti and L.~Vanzo,
  %``Rotating topological black holes,''
  Phys.\ Rev.\ D {\bf 57} (1998) 6127
   [Erratum-ibid.\ D {\bf 60} (1999) 109902]
  [gr-qc/9710123].
  %%CITATION = GR-QC/9710123;%%

%\cite{Gnecchi:2013mja}
\bibitem{Gnecchi:2013mja}
  A.~Gnecchi, K.~Hristov, D.~Klemm, C.~Toldo and O.~Vaughan,
  %``Rotating black holes in 4d gauged supergravity,''
  arXiv:1311.1795 [hep-th].

%\cite{Caldarelli:2008pz}
\bibitem{Caldarelli:2008pz}
  M.~M.~Caldarelli, R.~Emparan and M.~J.~Rodr\'{\i}guez,
  %``Black Rings in (Anti)-deSitter space,''
  JHEP {\bf 0811} (2008) 011
  [arXiv:0806.1954 [hep-th]].
  %%CITATION = ARXIV:0806.1954;%%

%\cite{Caldarelli:2012cm}
\bibitem{Caldarelli:2012cm}
  M.~M.~Caldarelli, R.~G.~Leigh, A.~C.~Petkou, P.~M.~Petropoulos, V.~Pozzoli and K.~Siampos,
  %``Vorticity in holographic fluids,''
  PoS CORFU {\bf 2011} (2011) 076
  [arXiv:1206.4351 [hep-th]].
  %%CITATION = ARXIV:1206.4351;%%

%\cite{Carter:1968ks}
\bibitem{Carter:1968ks}
  B.~Carter,
  %``Hamilton-Jacobi and Schr\"odinger separable solutions of Einstein's equations,''
  Commun.\ Math.\ Phys.\  {\bf 10} (1968) 280.
  %%CITATION = CMPHA,10,280;%%

\bibitem{Plebanski:1975}
  J.~F.~Pleba\'nski,
  %``A class of solutions of Einstein-Maxwell equations,''
  Annals Phys.\  {\bf 90} (1975) 196.

%\cite{Chow:2013gba}
\bibitem{Chow:2013gba}
  D.~D.~K.~Chow and G.~Comp\`ere,
  %``Dyonic AdS black holes in maximal gauged supergravity,''
  arXiv:1311.1204 [hep-th].
  %%CITATION = ARXIV:1311.1204;%%

%\cite{AlonsoAlberca:2000cs}
\bibitem{AlonsoAlberca:2000cs}
  N.~Alonso-Alberca, P.~Meessen and T.~Ort\'{\i}n,
  %``Supersymmetry of topological Kerr-Newman-Taub-NUT-AdS space-times,''
  Class.\ Quant.\ Grav.\  {\bf 17} (2000) 2783
  [hep-th/0003071];
  %%CITATION = HEP-TH/0003071;%%
%\cite{Klemm:2013eca}
%\bibitem{Klemm:2013eca}
  D.~Klemm and M.~Nozawa,
  %``Supersymmetry of the C-metric and the general Plebanski-Demianski solution,''
  JHEP {\bf 1305} (2013) 123
  [arXiv:1303.3119 [hep-th]].
  %%CITATION = ARXIV:1303.3119;%%

%\cite{Martelli:2013aqa}
\bibitem{Martelli:2013aqa}
  D.~Martelli and A.~Passias,
  %``The gravity dual of supersymmetric gauge theories on a two-parameter deformed three-sphere,''
  Nucl.\ Phys.\ B {\bf 877} (2013) 51
  [arXiv:1306.3893 [hep-th]].
  %%CITATION = ARXIV:1306.3893;%%

\end{thebibliography}
\end{document}